\documentclass[showpacs,twocolumn,pre]{revtex4}
\usepackage{amssymb}
\usepackage{graphicx}
\usepackage{dcolumn}
\usepackage{bm}

\def\be{\begin{equation}}
\def\ee{\end{equation}}
\def\bea{\begin{eqnarray}}
\def\eea{\end{eqnarray}}
\def\ba{\begin{array}}
\def\ea{\end{array}}

\def\b{\beta}
\def\c{\gamma}

\def\D{\Delta}

\def\vep{\varepsilon}
\def\k{\kappa}
\def\x{\xi}

\def\0{$\Gamma_0$}
\def\o{\omega}
\def\p{\phi}

\def\s{\sigma}

\def\l{\lambda}

\begin{document}

\title{Nanowire waveguide made from extremely anisotropic metamaterials}
\author{Y. J. Huang}
\author{W. T. Lu}
\email{w.lu@neu.edu}
\author{S. Sridhar}
\affiliation{Department of Physics and Electronic Materials Research Institute,
Northeastern University, Boston, Massachusetts 02115, USA}
\date{\today }

\begin{abstract}
Exact solutions are obtained for all the modes of wave propagation along 
an anisotropic cylindrical waveguide. Closed-form expressions for the energy flow on the
waveguide are also derived. For extremely anisotropic waveguide where 
the transverse permittivity is negative ($\vep_\bot<0$)
while the longitudinal permittivity is positive ($\vep_{||}>0$), 
only transverse magnetic (TM) and hybrid modes will propagate on the waveguide. 
At any given frequency the waveguide supports an infinite number of eigenmodes. 
Among the TM modes, at most only one mode is forward wave. 
The rest of them are backward waves which can have very large effective index. 
At a critical radius, the waveguide supports degenerate forward- and backward-wave modes 
with zero group velocity. These waveguides can be used as phase shifters and filters, 
and as optical buffers to slow down and trap light. 
\end{abstract}

\maketitle

\section{Introduction}

Since the realization of negative refraction \cite{Veselago} in microwaves \cite{Shelby}, 
there is renewed and intense interest in electromagnetic metamaterials. 
Negative refraction has added a new arena to physics, 
leading to new concepts such as perfect lens \cite{Pendry00,Lu05},
superlens \cite{Pendry00,Lu03,Fang}, and focusing by plano-concave lens \cite{Vodo05,Vodo06}.
Negative refraction has subsequently been realized in microwaves 
\cite{Parazzoli,Parimi04,Parimi03,Cubukcu,LuZ}, 
THz waves, and optical wavelengths \cite{Berrier,Dolling,Soukoulis07}, 
in metamaterials made of wire and split-ring resonators \cite{Smith04}
or photonic crystals \cite{Notomi,Gralak,Luo02}.

Metamaterials are artifically fabricated structures possessing
certain desirable properties which are not available in natural materials.
Metamaterials can have double negative index \cite{Shalaev} or single negative index. 
Metamaterials can be periodic, such as photonic crystals \cite{Joannopoulos}. 
They can also be non-periodic, 
such as the materials for cloaking \cite{Schurig06}.
They can also be made to be anisotropic and have indefinite index \cite{Smith03,Hoffman,Lu07}. 
Indefinite index matematerials can be used to make hyperlens \cite{LiuZ,Smolyaninov}. 
This range of properties opens an infinite possibilities to use matematerials
in frequencies from microwave all the way up to the visible.

Wave propagation in waveguide of nanometer size \cite{Takahara} has unique properties.
In this paper, we consider wave propagation along anisotropic nanowires.
In the case where the transverse permittivity is negative 
while the longitudinal one is positive ($\vep_\bot<0$, $\vep_{||}>0$), 
these indefinite index waveguides can support both forward and backward waves.
High effective index can be obtained for these modes. 
These waveguides can also support degenerate modes
which can be used to slow down and trap light.

In Sec. II, we derive the formulas for all the modes on the anisotropic cylinders. 
Exact solutions for all the modes and closed-form expressions 
for the energy flow will be obtained.
Possible zero net-energy flow modes will also be discussed.
The situation for trapping light is presented in Sec. III.
In Sec. IV, we propose the realization of nanowires made of indefinite index medium, which
is confirmed in finite-difference time-domain simulations. 
We conclude in Sec. V with possible applications for these anisotropic nanowires.

\section{Wave propagation and energy flow on anisotropic cylindrical waveguides}

\begin{figure}[htbp]
\center{
\includegraphics [angle=0, width=7cm]{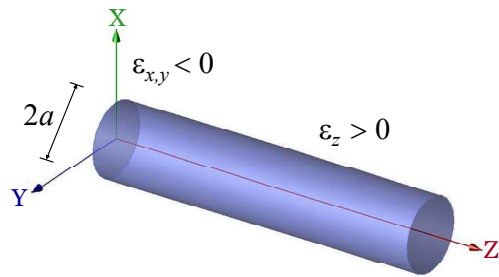}}
\caption{(Color online) An anisotropic cylindrical waveguide with axis along the $z$-axis. 
The longitudinal permittivity is
positive while the transverse permittivity is negative.}
\label{fig-sketch}
\end{figure}

We consider wave propagation on a cylindrical waveguide. 
The axis of the waveguide is along the $z$-direction as shown in Fig. \ref{fig-sketch}.
The waveguide is nonmagnetic and has an anisotropic optical property
\be
\vep_x=\vep_y=\vep_t\neq\vep_z.
\ee

The waves will propagation along the cylinder axis with
\be
{\bf E}={\bf E}_0e^{i(\b z-\o t)},\quad {\bf H}={\bf H}_0e^{i(\b z-\o t)}.
\ee
Here $\b$ is the propagation wave number along the waveguide.

Due to the symmetry of the waveguide, all the field components can be expressed 
in terms of the longitudinal components $E_z$ and $H_z$.
In the polar coordinate system, one has for the fields inside the waveguide $r<a$ with
$a$ the radius,
\bea
E_r&=&i{1\over \vep_tk_0^2-\b^2}\Big(\b\partial_r E_z+k_0{1\over r}\partial_\p H_z\Big),\nonumber \\
E_\p&=&i{1\over \vep_tk_0^2-\b^2}\Big(\b{1\over r}\partial_\p E_z-k_0\partial_r H_z\Big),\nonumber \\
H_r&=&i{1\over \vep_tk_0^2-\b^2}\Big(-\vep_tk_0{1\over r}\partial_\p E_z
+\b\partial_r H_z\Big),\nonumber \\
H_\p&=&i{1\over \vep_tk_0^2-\b^2}\Big(\vep_tk_0\partial_r E_z+\b{1\over r}\partial_\p H_z\Big).
\eea
Here $k_0$ is the wave number in the vacuum.

The wave equations for the longitudinal components inside the waveguide are
\bea
&(\partial^2_x+\partial^2_y)E_z+\vep_z(k_0^2-\b^2/\vep_t)E_z=0,&\nonumber \\
&(\partial^2_x+\partial^2_y)H_z+(\vep_tk_0^2-\b^2)H_z=0.&
\eea
The waveguide is free-standing in air, so the wave equations for $r>a$ are 
given by the above equations with
the permittivity replaced by unity.
The solutions are expressed in terms of the Bessel functions of various kinds
\bea
E_z&=&AJ_n(Kr)e^{in\p},\ \quad r<a, \nonumber \\
&=&CK_n(\kappa_0 r)e^{in\p},\quad r>a \label{eq-Ez}
\eea
and
\bea
H_z&=&BI_n(\kappa r)e^{in\p}, \qquad r<a, \nonumber \\
&=&DK_n(\kappa_0 r)e^{in\p},\quad r>a.\label{eq-Hz}
\eea
The coefficients will be determined by matching the boundary conditions. Here
\bea
K&=&\sqrt{\vep_z}\sqrt{k_0^2-\b^2/\vep_t},\nonumber \\
\kappa&=&\sqrt{\b^2-\vep_tk_0^2},\nonumber \\
\kappa_0&=&\sqrt{\b^2-k_0^2}.
\eea
We only consider the extremely anisotropic case such that the longitudinal permittivity is
positive while the transverse permittivity is negative
\be
\vep_z>0, \quad \vep_t<0.
\ee
One can see that due to the anisotropic nature of the waveguide, $H_z$ and $E_z$ inside
the waveguide will have completely different behaviors.

The continuity of $E_z$ and $H_z$ at the interface $r=a$ gives
\be
{C\over A}={J_n(Ka)\over K_n(\kappa_0 a)},\quad
{D\over B}={I_n(\k a)\over K_n(\kappa_0 a)}.\label{eq-ABCD}
\ee
The continuity of $E_\p$ at the interface gives
\be
in\b{\k_0^2-\k^2\over k_0a^2\k_0^2\k^2}={B\over A}{I_n(\k a)\over J_n(Ka)}
\Big[g_n(\k_0a)+h_n(\k a)\Big]
\ee
with the following defined functions
\bea
g_n(x)&=&-{K'_n(x)\over xK_n(x)}={K_{n-1}(x)\over xK_n(x)}+{n\over x^2},\nonumber \\
h_n(x)&=&{I'_n(x)\over xI_n(x)}={I_{n-1}(x)\over xI_n(x)}-{n\over x^2}.
\eea
The continuity of $H_\p$ at the interface gives
\be
in\b{\k^2-\k_0^2\over k_0a^2\k_0^2\k^2}={A\over B}{J_n(Ka)\over I_n(\k a)}
\Big[g_n(\k_0a)-\vep_zf_n(Ka)\Big]
\ee
with
\be
f_n(x)={J'_n(x)\over xJ_n(x)}={J_{n-1}(x)\over xJ_n(x)}-{n\over x^2}.
\ee
Thus we obtain the equation for all the modes
\bea
&&\Big[g_n(\kappa_0 a)-\vep_zf_n(K a)
\Big]
\Big[g_n(\kappa_0 a)+h_n(\kappa a)
\Big]\nonumber \\
&&=n^2\Big[(\k_0a)^{-2}-(\k a)^{-2}\Big]
\Big[(\k_0a)^{-2}-\vep_t(\k a)^{-2}\Big].
\label{mode-eqn-all}
\eea

For wave propagation on the cylindrical waveguide, 
the components of the Poynting vector are
\bea
S_z&=&{1\over 4\pi}(E_rH_\p^*-E_\p H_r^*),\nonumber \\
S_r&=&-{1\over 4\pi}(E_zH_\p^*-E_\p H_z^*),\nonumber \\
S_\p&=&{1\over 4\pi}(E_zH_r^*-E_rH_z^*).
\eea
The physical Poynting vector is given by $\Re {\bf S}$.

The total energy flow along the waveguide is the sum of energy flow inside and outside the
waveguide
\be
P_z=P^{\rm in}_z+P^{\rm out}_z
\ee
with
\be
P^{\rm in}_z=2\pi\int_0^a S_zrdr,\quad P^{\rm out}_z=2\pi\int_a^\infty S_zrdr.
\ee
Following Ref. \cite{Tsakmakidis06}, the total energy flow is normalized as
\be
\langle P_z\rangle={P^{\rm in}_z+P^{\rm out}_z \over |P^{\rm in}_z|+|P^{\rm out}_z|}.
\ee
Thus one has $-1<\langle P_z\rangle<1$.

In the following, we discuss different modes in detail.

\subsection{TE modes}

For the transverse electric (TE) modes, $E_z=0$. 
The longitudinal magnetic field is
given by Eq. (\ref{eq-Hz}). One has
\bea
E_\p&=&i{k_0\over \k^2}\partial_r H_z=i{k_0\over \k}BI_1(\kappa r),
\qquad r<a,\nonumber \\
&=&i{k_0\over \k_0^2}\partial_r H_z=-i{k_0\over \k_0}DK_1(\kappa r),
\quad r>a.
\eea
The continuity of $E_\p$ at the interface requires that
\be
h_0(\kappa a)+g_0(\kappa_0 a)=0.
\ee
For materials without loss, each term on the left side is positive, 
thus there is no solution. 
The waveguide does not support TE modes. 
This is exactly like that of a metallic wire which does not
support TE surface waves since current must flow along the waveguide.

Only when $\vep_t>1$, the waveguide will support TE modes, like an 
ordinary dielectric fiber.

\subsection{TM modes}
For the transverse magnetic (TM) modes, $H_z=0$. The longitudinal electric field is
given by Eq. (\ref{eq-Ez}). One has
\bea
H_\p&=&-i\vep_z{k_0\over K}AJ_1(Kr),\quad r<a,\nonumber \\
&=&i{k_0\over \k_0}CK_1(\k_0r),\qquad r>a.
\eea
The continuity of $H_\p$ leads to the equation
\be
\vep_zf_0(K a)=g_0(\kappa_0 a).
\label{eq-TM}
\ee
The solutions to this equation give all the TM modes.

\subsubsection{Band structure of TM modes}
We first consider the solutions for fixed and real values of $\vep_z$ and $\vep_t$. 
This is normally associated with a fixed $k_0$. 
It is convenient to consider solution in the form of $Ka$ or the reduced radius $k_0a$
as a function of $\k_0a$.
The wave  number along the waveguide can be obtained through
$\b=(k_0^2+\k_0^2)^{1/2}$.
Before we seek general solutions, it is better
to consider the solutions in certain limits
to reveal some important features of the TM modes on the anisotropic waveguide.

For the TM  modes close to the light line, $\k_0a\to0$, one has
\[
Ka\simeq x_{0,m}-{\vep_z\over x_{0,m}}(\k_0a)^2(\ln{\k_0a\over 2}+\c).
\]
Here we have used $K_0(x)=-\ln(x/2)-\c$ for small augument with $\c$ the Euler constant.
For complex $\vep_t$ with $\Re\vep_t<0$ and $\Im\vep_t>0$, 
the real and imaginary parts of $\b$ of the allowed modes will have the same signs.
These modes are forward waves, similar to that of an ordinary optical fiber.
We note that close to the light line, the property of the TM modes 
of the anisotropic waveguide is similar to that of an isotropic fiber with
$\vep=1+\vep_z(1-\vep_t^{-1})$.
 
In the limit of long wavelength or small waveguide radius, $k_0a\ll 1$, 
Eq. (\ref{eq-TM}) is reduced to
\be
\vep_zf_0(\k_0a/\eta)=g_0(\k_0a) \label{eq-TM-longwave}
\ee
with $\eta=\sqrt{-\vep_t/\vep_z}$.
This equation gives an infinite number of solutions 
 $\k_0a=\x_{0,m}$ with $m=1,2,3,\cdots$. 
This indicates that the anisotropic waveguide supports infinite number of propagating modes,
no matter how thin the waveguide is.
For $\k_0a\to\infty$, since $g_0(\k_0a)\simeq (\k_0a)^{-1}\to0$, 
one has $\x_m\simeq \eta x_{1,m}$.
Here $x_{n,m}$ is the $m$-th zero of $J_n(x)$ away from the origin.
For the $m$-th TM band, one has $0\leq \k_0a\leq \x_{0,m}$.
The $m$-th band starts with 
$k_0a=x_{0,m}/\sqrt{\vep_z-\vep_z/\vep_t}$ when $\k_0a=0$ and
ends at $k_0a=0$ when $\k_0a=\x_{0,m}$.
The modes with $\k_0\gg k_0$ 
have $d(k_0a)/d(\k_0a)<0$ and are backward wave. 
It will be obvious if we include small imaginary part in $\vep_t$ with $\Im\vep_t>0$.
The equation will give $\b$ with the real and imaginary parts having opposite signs.
The energy flow is opposite to the phase velocity,
which will be discussed later.

\begin{figure}[htbp]
\center{
\includegraphics [angle=0, width=7.5cm]{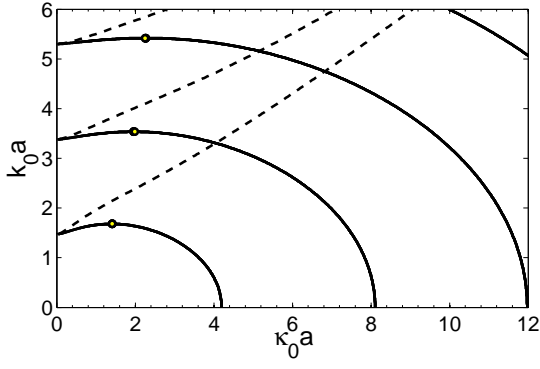}}
\caption{Band structure of the guided TM modes on an anisotropic waveguide 
of radius $a$ with $\vep_t=-3$ and $\vep_z=2$. Open circles denote the 
degenerate points of forward-wave and backward-wave modes. The dashed lines are for
a dielectric waveguide with $\vep=1+\vep_z(1-\vep_t^{-1})$.}
\label{fig-band}
\end{figure}

\begin{figure}[htbp]
\center{
\includegraphics [angle=0, width=7.5cm]{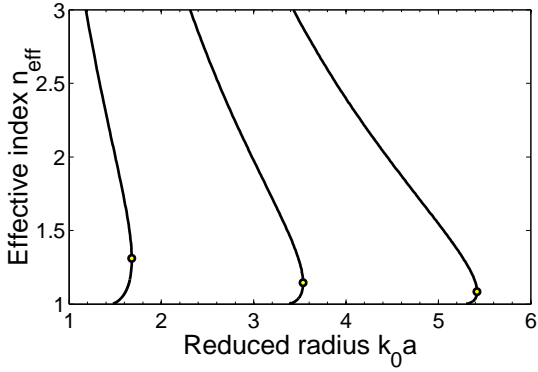}}
\caption{Effective index $n_{\rm eff}$ as a function of the reduced radius
$k_0a$ for the first three bands of TM modes on an anisotropic waveguide 
with $\vep_t=-3$ and $\vep_z=2$. Open circles denote the 
degenerate points of forward-wave and backward-wave modes.}
\label{fig-neff}
\end{figure}

For arbitrary values of $\k_0a$, the solution must be sought numerically.
Since the right-hand side of Eq. (\ref{eq-TM}) is always positive, 
the solution requires that $J_1(Ka)$ and $J_0(Ka)$ have different signs. 
For the $m$-th band, since $0\leq \k_0a\leq \x_{0,m}$ with $\x_{0,m}$
the solutions of Eq. (\ref{eq-TM-longwave}),
one has $x_{0,m}\leq Ka\leq \x_{0,m}/\eta<x_{1,m}$. 
For each $\k_0a$ value, the $Ka$ value can be searched within
$[x_{0,m},x_{1,m}]$ to satisfy Eq. (\ref{eq-TM}). 
Once the corresponding $Ka$ is found, the reduced radius can be obtained as
\[
k_0a={\sqrt{(-\vep_t/\vep_z)(Ka)^2-(\k_0a)^2}\over \sqrt{1-\vep_t}}.
\]
For the $m$-th band, the corresponding transverse electric field $E_z$ will have $m$ nodes.
The band structure for a waveguide with $\vep_t=-3$ and $\vep_z=2$
is shown in Fig. \ref{fig-band}. 
The effective index of the waveguide $n_{\rm eff}\equiv \b/k_0$ 
is also evaluated and plotted in Fig. \ref{fig-neff}.

Unlike an ordinary fiber where for each band, $d(k_0a)/d(\k_0a)>0$, 
each TM band of the anisotropic waveguide starts with $d(k_0a)/d(\k_0a)>0$ 
for small $\k_0a$ or near the light line. 
At certain value of $\k_0a$ or $k_0a$ which is marked in Fig. \ref{fig-band}, 
$d(k_0a)/d(\k_0a)=0$. 
Further increasing $\k_0a$ results in $d(k_0a)/d(\k_0a)<0$. 
The band ends at a finite $\k_0a=\x_{0,m}$.
Immediately below the point $k_0a$ where $d(k_0a)/d(\k_0a)=0$, 
each band has two modes with opposite signs of $d(k_0a)/d(\k_0a)$. 
One mode is forward wave and the other backward wave. 
This will be discussed later in the paper.

\begin{figure}[htbp]
\center{
\includegraphics [angle=0, width=7.5cm]{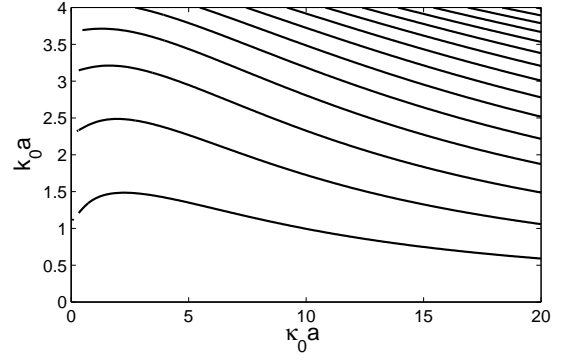}}
\caption{Band structure of the guided TM modes on an anisotropic waveguide 
of radius $a$ with $\vep_t$ and  $\vep_z$ given by Eq. (\ref{ep-metal-dielec}) 
with $a=10/k_p$ and $\vep_a=2.25$.}
\label{fig-drude-band}
\end{figure}

\begin{figure}[htbp]
\center{
\includegraphics [angle=0, width=8cm]{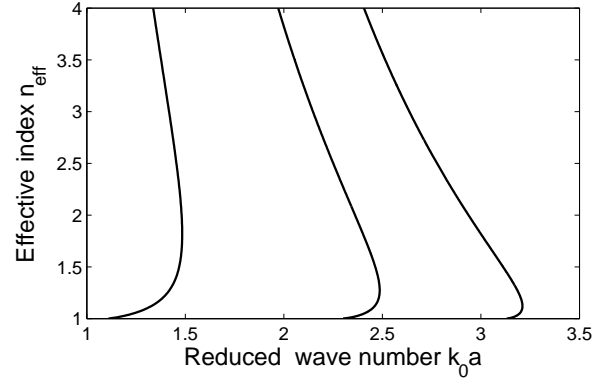}}
\caption{Effective index $n_{\rm eff}$ as a function of the reduced wave number
$k_0a$ for the first three bands of TM modes on an anisotropic waveguide with 
$\vep_t$ and  $\vep_z$ given by Eq. (\ref{ep-metal-dielec}) 
with $a=10/k_p$ and $\vep_a=2.25$.}
\label{fig-drude-neff}
\end{figure}

We next consider a waveguide of a fixed radius $a$ 
with the following permittivity 
\bea
\vep_t&=&{1\over 2}(1+\vep_a-k_p^2/k_0^2),\nonumber \\
\vep_z&=&2\vep_a(k_0^2-k_p^2)/[k_0^2(1+\vep_a)-k_p^2].
\label{ep-metal-dielec}
\eea
Here $\vep_a$ and $k_p$ are positive constants. 
The realization of this property will be discussed later in Sec. IV.
If $k_0<k_p/(1+\vep_a)^{1/2}$, one has $\vep_t<0$ and $\vep_z>0$.
The band structure of the TM modes on this waveguide is obtained 
by numeric means and plotted in  Fig. \ref{fig-drude-band} 
with the corresponding effective index in Fig. \ref{fig-drude-neff}.
For this waveguide, there is no cutoff of $\k_0a$ for each band.
This is because that as $k_0\to 0$, $\vep_z\simeq 2\vep_a$ and $\vep_t\to-\infty$, thus
$\eta=\sqrt{-\vep_t/\vep_z}\to\infty$. 
The cutoff $\x_{0,m}\simeq \eta x_{1,m}\to\infty$.

We point out that the Pad\'e approximant for the function $f_0(x)\equiv -x^{-1}J_1(x)/J_0(x)$ 
can be used to obtain good estimate of the solutions.
This will be discussed in the Appendix.

If $\vep_t>0$ and $\vep_z<0$, the waveguide also supports TM modes.
The details will not be presented here.

\subsubsection{Energy flow of TM modes}
Within the waveguide, one has $E_z=AJ_0(Kr)$,
$E_r=i(\vep_z\b/\vep_t  K)AJ_1(Kr)$,
and $H_\p=i(\vep_zk_0/K)AJ_1(Kr)$.
So the Poynting vector component along the axis of the waveguide is
\be
S_z={\b\over \vep_tk_0a^2} \Big|{J_1(Kr)\over J_1(Ka)}\Big|^2.
\ee
Here we set the coefficient $A=K/[\vep_zk_0aJ_1(Ka)]$. 
Since $\vep_t<0$, the energy flow inside the nanowire is always opposite to the
phase velocity.

For the field in the air $r>a$, one has
$E_z=CK_0(\k_0r)$,
$E_r=i(\b\/\k_0)CK_1(\k_0r)$, and
$H_\p=i(k_0/\k_0)CK_1(\k_0r)$,
thus
\be
S_z={\b\over k_0a^2}\Big|{K_1(\k_0r)\over K_1(\k_0a)}\Big|^2.
\ee
Here the coefficient $C=-\k_0/[k_0aK_1(\k_0a)]$.

For the TM modes, one has
\bea
P^{\rm in}_z&=&{\b\over 2\vep_tk_0a^2}\int_0^a\Big|{J_1(Kr)\over J_1(Ka)}\Big|^2rdr,
\nonumber \\
P^{\rm out}_z&=&{\b\over 2k_0a^2}\int_a^\infty\Big|{K_1(\k_0r)\over K_1(\k_0a)}\Big|^2rdr.
\eea
The above integrals can be carried out and more compact expressions for the energy flow
can be obtained as
\bea
P^{\rm in}_z&=&{\b \over 4\vep_tk_0(Ka)^2}\Big[{1\over f_0^2(Ka)}
+{2\over f_0(Ka)}+(Ka)^2\Big]\nonumber \\
&=&-{\b\over 4k_0\vep_z^2 f^2_0(Ka)}{\vep_z^2f'_0(Ka)\over \vep_t Ka},\nonumber \\
\nonumber \\
P^{\rm out}_z&=&{\b \over 4k_0(\k_0a)^2}\Big[{1\over g_0^2(\k_0a)}
+{2\over g_0(\k_0a)}-(\k_0a)^2\Big]\nonumber \\
&=&-{\b \over 4k_0g_0^2(\k_0a)}{g'_0(\k_0a)\over \k_0a}.
\eea
Here $g'_0(x)$ and $f'_0(x)$ are the derivatives of $g_0(x)$ and $f_0(x)$,
respectively.

\begin{figure}[htbp]
\center{
\includegraphics [angle=0, width=8.2cm]{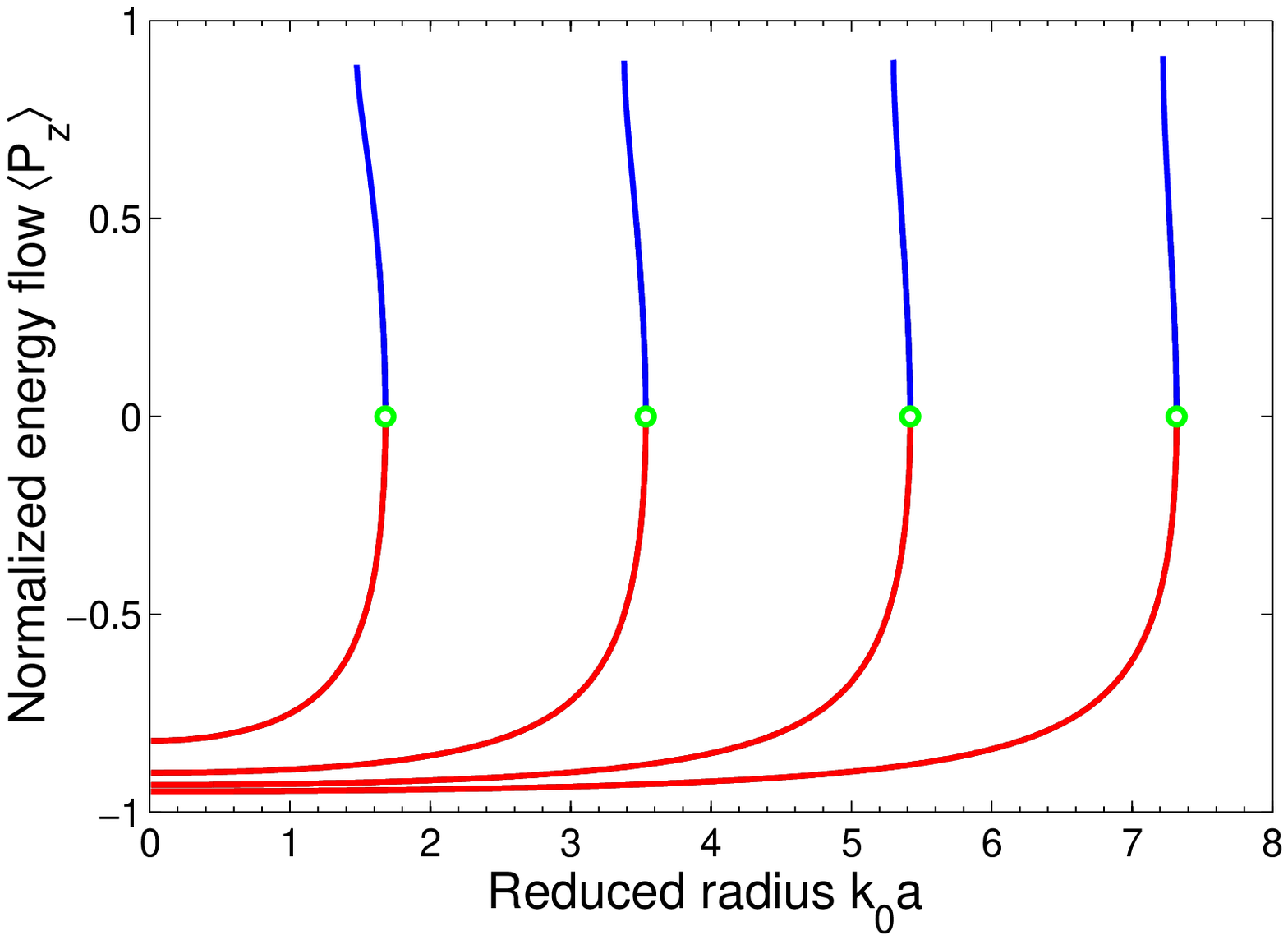}}
\caption{(Color online) Normalized energy flow $\langle P_z\rangle$ 
for the first four bands of the TM modes on
an anisotropic waveguide with $\vep_t=-3$ and $\vep_z=2$. Here we set $\b>0$.
Open circles denote the degenerate points where
$\langle P_z\rangle=0$.}
\label{fig-flow-TM}
\end{figure}

For convenience, we set $\b>0$ throughout the paper. 
Since $g'_0(x)<0$ and $f'_0(x)<0$, one has $P^{\rm in}_z<0$ 
and $P^{\rm out}_z>0$.
In this convention, if $\langle P_z\rangle>0$, this indicates that 
the energy flow and the phase propagation are in the same directions and
the mode is a forward-wave mode. Otherwise $\langle P_z\rangle<0$, 
the group velocity and the phase velocity 
are in the opposite direction and the mode is a backward-wave mode.
The normalized energy flow for TM  modes on a waveguide with $\vep_t=-3$ and $\vep_z=2$
is shown in Fig. \ref{fig-flow-TM}.
We note that for some portion of the bands the value of $\langle P_z\rangle$ 
is negative and thus these modes are backward waves.

\subsubsection{Forward-wave and backward-wave TM modes}

There are three ways to determine whether a mode is a forward wave or backward wave.
One is through the sign of the derivative $d(k_0a)/d(\k_0a)$. From the band structure,
we have already notice that for the modes near the light line, $d(k_0a)/d(\k_0a)>0$.
These modes are forward waves. For large $\k_0a$ or small $k_0a$, one has 
$d(k_0a)/d(\k_0a)<0$, these modes are backward waves. From the band structure shown
in Fig. \ref{fig-band}, the majority of the modes are backward waves.

The second way is through the sign of $\langle P_z\rangle$.
For the TM modes when $\k_0a\to\infty$, one has $f_0(Ka)\to0$ since $g_0(\k_0a)\to 0$.
One thus has $Ka\to x_{0,m}$. This solution leads to the divergence of $P^{\rm in}_z$ 
which is negative and
the vanishing of $P^{\rm out}_z$ which is positive, subsequently $\langle P_z\rangle\to-1$, 
these modes are all backward waves. Correspondingly, one has
$d(k_0a)/d(\k_0a)\to-\infty$ for $\k_0a\to\infty$. This  is evident from the band structure
shown in Fig. \ref{fig-band}.

In the following, we prove that $d(k_0a)/d(\k_0a)\geq 0$ leads to 
$\langle P_z\rangle\geq0$ and vice versa.
We consider the derivative
\[
{d(Ka)\over d(\k_0a)}=-{\vep_z \over \vep_t Ka}\Big[\k_0a
+(1-\vep_t)k_0a{d(k_0a)\over d(\k_0a)}\Big].
\]
Thus one has
\[
\quad {d(Ka)\over d(\k_0a)}+{\vep_z \k_0a\over \vep_t Ka}\geq0 \quad 
{\rm if}\quad {d(k_0a)\over d(\k_0a)}\geq0.
\]
On the other hand, according to the eigen equation $\vep_zf_0(Ka)=g_0(\k_0a)$, one has
\[
{d(Ka)\over d(\k_0a)}={g'_0(\k_0a)\over \vep_zf'_0(Ka)}.
\]
Making use of the above expression, one arrives at the inequality
\[
-{\vep_z^2 f'_0(Ka)\over \vep_t Ka}-{g'_0(\k_0a)\over \k_0a}\geq0,
\]
and subsequently
\be
P^{\rm in}_z+P^{\rm out}_z\geq0, \quad {\rm if}\quad {d(k_0a)\over d(\k_0a)}\geq 0.
\ee
Similarly one has
\be
P^{\rm in}_z+P^{\rm out}_z\leq0 \quad {\rm if}\quad {d(k_0a)\over d(\k_0a)}\leq 0.
\ee
So the condition for $P_z=0$ can be allocated from the band structures as
shown in Fig. \ref{fig-band}, \ref{fig-drude-band} when $d(k_0a)/d(\k_0a)=0$.

\begin{figure}[htbp]
\center{
\includegraphics [angle=0, width=8.2cm]{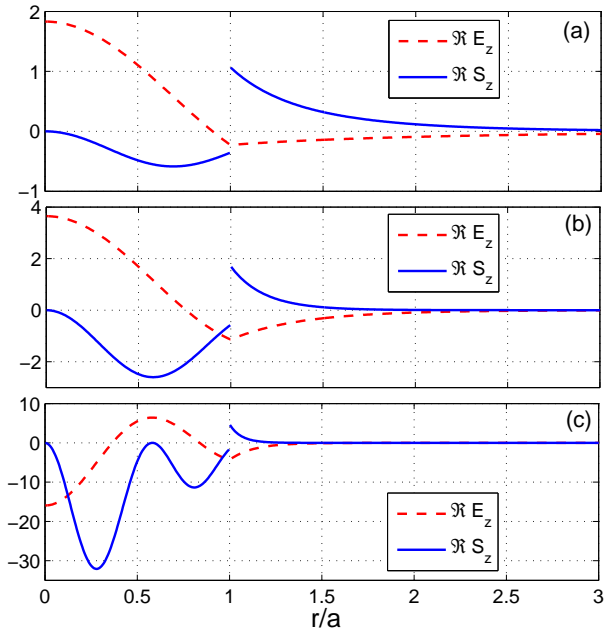}}
\caption{(Color online) The longitudinal electric field and Poynting vector for the first three TM mode 
on an anisotropic waveguide of radius $a$ with $\vep_t=-3+0.05i$ and $\vep_z=2$
at $k_0a=1.6$ with (a) $\b a=  1.7112+0.0067i$, (b) $\b a=  2.7250 -0.0397i$, (c) $\b a=  7.5756 - 0.0676i$.
The imaginary parts $\Im E_z$ and $\Im S_z$ are small and not plotted.}
\label{fig-field}
\end{figure}

The third way to determine whether a mode is a forward or backward wave
is through the relative sign of the real and imaginary parts of $\b$ if dissipation is included.
For example we consider $\vep_t=-3+0.05i$ and $\vep_z=2$. 
At $k_0a=1.6$, the wave numbers of the first three eigenmodes
are $\b a=\pm(1.7112+0.0067i),\pm(2.7250 -0.0397i),\pm(7.5756 -0.0676i)$. 
Since the free space wave length is $\l=3.927a$, 
this is a subwavelength waveguide.
For the TM modes, exept for the first mode, all the other modes are backward-wave modes
since for those modes $\Re\b$ and $\Im \b$ have different signs.
The normalized energy flow is
$\langle P_z\rangle= 0.5151 - 0.0020i,  -0.4002 - 0.0058i,  -0.8760 - 0.0078$ 
for the above three modes, respectively. Here we set $\Re\b>0$.
The field and Poynting vector profiles are plotted in Fig. \ref{fig-field}.

There is an interesting feature of the modes on the anisotropic waveguide.
At a fixed $k_0$, for $a<a_m\equiv\sqrt{-\vep_t/\vep_z(1-\vep_t)}x_{0,m}/k_0$, 
the $m$-th band TM modes are backward waves. 
If the radius $a>a_m$, the waveguide supports two TM modes for the $m$-th band,
one forward and one backward. 
At $a=a_c$, these two modes become degenerate and the total energy flow is zero.  
This can be seen in Fig. \ref{fig-band}, \ref{fig-neff}, \ref{fig-flow-TM} 
where degenerate points are marked.
Further increasing the radius, the waveguide
will no longer support the $m$-th band. 
The critical radius $a_c$ is located such that $\langle P_z\rangle=0$,
$d(k_0a)/d(\k_0a)=0$ or $dn_{\rm eff}/da=\infty$.
These degenerate modes can be used to slow down and even trap light.
This will be discussed in the next section.

\subsection{Hybrid modes}
The modes with both $E_z\neq 0$ and $H_z\neq 0$ are called hybrid modes.
Their dispersions are contained in the solutions of Eq. (\ref{mode-eqn-all}) with $n\neq 0$.
We recast the equation in the following form
\be
\vep_zf_n(K a)
=g_n(\x)-{n^2(\x^{-2}-y^{-2})(\x^{-2}-\vep_t y^{-2})\over
g_n(\x)+h_n(y)}.
\label{eq-hybrid}
\ee
Here we use the notation $\x=\k_0a$ and $y=\k a$.
Note that $Ka=y/\eta$ with $\eta=\sqrt{-\vep_t/\vep_z}$.

\subsubsection{Band structure of hybrid modes}
At a fixed wavelength $\l$ or wave number $k_0$, $\vep_t$, $\vep_z$ are constant. 
Since $Ka=y/\eta$, 
if we use $\x=\k_0a$ as a free parameter, the eigen
equation gives a single value of $Ka$ or $y$ for each $\x$.
Since $y=\sqrt{\x^2+(1-\vep_t)(k_0a)^2}$, so the eigen equation actually gives
the reduced radius $k_0a$ for each $\xi=\k_0a$.

In the limit of long wavelength or small waveguide radius, $k_0a\to 0$, 
Eq. (\ref{eq-hybrid}) is reduced to
\be
\vep_zf_n(\x/\eta)=g_n(\x).\label{eq-hybrid-longwave}
\ee
This will give a discrete set of solutions $\k_0a=\xi_{n,m}$ for each $n$.
For $\eta$ and $\vep_z$ not very small, the values $\x_{n,m}$ can be obtained
approximately by making use of the asymptotic behavior $g_n(x)\sim x^{-1}$ for $x\to\infty$
and the Pad\'e approximant of $f_n(x)$, which will be discussed in the Appendix.
The anisotropic waveguide supports infinite number of hybrid modes,
no matter how thin the waveguide is.

Close to the light line, $\x\to0$, the eigen equation can be simplified.
We consider the hybrid modes with different $n$ separately.

For $n=1$, one has $g_1(x)\simeq x^{-2}-\ln(x/2)-\c$ for $x\to0$,
thus Eq. (\ref{eq-hybrid}) becomes
\be
\vep_zf_1(Ka)
\simeq -2\big(\ln {\x\over 2}+\c\big) +h_1(y)+{1+\vep_t\over y^2}.
\ee
with $\x\ll y$. One has the solution $Ka\to x_{1,m}$ for $\x\to0$.
Note that throughout this paper, we use $x_{n,m}$ to denote the $m$-th zero of $J_n(x)$.
Also that $x_{n,0}=0$ for $n\geq 1$. Since $x_{1,0}=0$, 
special care should be taken for solutions $Ka\sim 0$.
Making use of the asymptotic behaviors $h_n(x)\simeq nx^{-2}+1/[2(n+1)]$ 
and $f_n(x)\simeq nx^{-2}-1/[2(n+1)]$ for $x\to0$ with $n\geq 1$, 
one gets
\be
Ka\simeq {1\over \eta}\sqrt{1+\vep_t\over \ln(\x/2)+\c-(\vep_z+1)/8}.
\ee
From this expression, one can see that only if
 $\vep_t<-1$, there will be solution for $Ka\sim 0$ when $\x\to0$. 
The above expression gives the dispersion of the first band with $n=1$.
The $m$-th band will start with $Ka=x_{1,m-1}$ and end with $Ka=\x_{1,m}/\eta$.
One has $\x_{1,1}<\eta x_{1,1}$.
However if $-1<\vep_t<0$, there will be no solution for $Ka<x_{1,1}$.
All the allowed modes will have finite $Ka$.
For the $m$-th band, one has $x_{1,m}\leq Ka\leq \x_{1,m}/\eta$ 
with $\x_{1,1}>\eta x_{1,1}$.
The first band starts with $Ka=x_{1,1}$.
For any $x_{1,m}>0$, one has
\be
Ka\simeq x_{1,m}-{\vep_t x_{1,m}\over (\eta x_{1,m})^2[-2(\ln(\x/2)+\c) +h_1(\eta x_{1,m})]+\vep_t+1}.
\ee
The hybrid modes near the light line with $n=1$ are all forward waves.

For $n\geq 3$, one has the following asymptotic behavior of $g_n(x)$
for $x\to 0$,
\be
g_n(x)\simeq {n\over x^2}+{1\over 2(n-1)}+ax^2.\label{gn-small-z}
\ee
with $a=-1/[8(n-2)(n-1)^2]$.
The eigen equation (\ref{eq-hybrid}) is reduced to
\be
a^2{\x}^{6}+c_{2}\x^4+c_{1}\x^2+c_{0}=0 \label{eq-hybrid-small-z}
\ee
with%
\bea
c_{2} &=&a\Big({1\over n-1}+h_{n}-\varepsilon _{z}f_{n}\Big), \nonumber \\
c_{1} &=&2na+{1\over 4(n-1)^2}+{h_{n}-\varepsilon _{z}f_{n}\over 2(n-1)}
-\varepsilon_{z}f_{n}h_{n}-n^{2}{\frac{\varepsilon _{t}}{y^{4}}}, \nonumber \\
c_{0} &=&{n\over n-1}+n(h_n-\vep_zf_n)+{n^2(1+\vep_t)\over y^2}. \label{eq-hybrid-coeff}
\eea
The eigen modes on the light line are given by the equation $c_0=0$.
The exsitence of solution requires that $c_0\geq 0$.
Only the positive root with small magnitude of the above cubic polynomial will give
the dispersion for the modes near the light line. 
Since $\x$ is small, the physical solution can be approximated
as $\x^2=-c_0/c_1-c_2c_0^2/c_1^3-a^2c_0^3/c_1^4$.
Those modes all have $Ka>x_{n,m}$ for the $m$-th band.

For $n=2$, the asymptotic behavior of $g_n(x)$ is still given by 
Eq. (\ref{gn-small-z}), but with a non-constant coefficient $a=[\ln(x/2)+\c]/4$.
The eigen equation near the light line is reduced further 
from Eq. (\ref{eq-hybrid-small-z}) to
\be
(\k_0a)^2\big(\ln{\k_0a\over 2}+\c\big)+c_0=0
\ee
with $c_0$ given in Eq. (\ref{eq-hybrid-coeff}) with $n=2$.

\begin{figure}[htbp]
\center{
\includegraphics [angle=0, width=7.6cm]{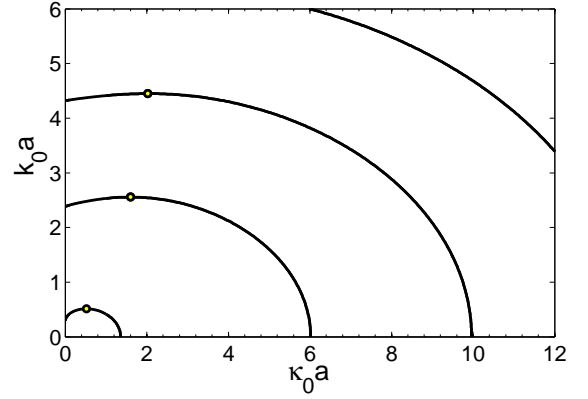}}
\caption{Band structure of the guided hybrid modes with $n=1$ on an anisotropic waveguide 
with $\vep_t=-3$ and $\vep_z=2$. Open circles denote the 
degeneracy of forward-wave and backward-wave modes.}
\label{fig-band-hybrid}
\end{figure}

\begin{figure}[htbp]
\center{
\includegraphics [angle=0, width=7.6cm]{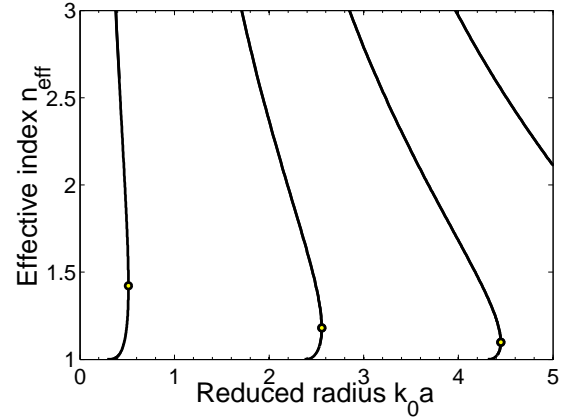}}
\caption{Effective index $n_{\rm eff}$ as a function of the reduced radius
$k_0a$ for the first three bands of hybrid modes with $n=1$ on an 
anisotropic waveguide with $\vep_t=-3$ and $\vep_z=2$. Open circles denote the 
degeneracy of forward-wave and backward-wave modes.}
\label{fig-neff-hybrid}
\end{figure}

For the allowed eigenmodes of the $m$-th hybrid band, 
one has the range $0\leq \k_0a<\x_{n,m}$
with $\x_{n,m}$ the solutions of Eq. (\ref{eq-hybrid-longwave}).
The solutions near both ends of the above range can be obtained 
analytically as we have done.
For arbitrary $\xi$ within this range, the solution must be obtained numerically.
However only when $\vep_t<-1$, one can have eigenmodes with $0<Ka<x_{n,1}$.
For the $m$-th band, one has $x_{n,m}<Ka\leq\x_{n,m}/\eta$. 
Otherwise, the solution for the first band requires 
$x_{n,1}<Ka\leq \x_{n,1}/\eta$ with $\x_{n,1}>\eta x_{n,1}$. 
The band structure for hybrid modes on a waveguide
with $\vep_t=-3$ and $\vep_z=2$
is shown in Fig. \ref{fig-band-hybrid}. 
The effective index of the waveguide $n_{\rm eff}\equiv \b/k_0$ 
is also evaluated and plotted in Fig. \ref{fig-neff-hybrid}.

\subsubsection{Energy flow of hybrid modes}

The energy flow can also be evaluated for hybrid modes. The expression for $S_z$ is 
much more complex than that of the TM modes. However the final
expression for $P_z$ is much simpler than expected, once the integrals are all carried out.
One has
\bea
P_z^{\rm in} &=&{\b\over 4k_0}\Bigg\{-{\vep_z^2\over \vep_tKa}(g_n+h_n)f'_n 
-{1\over \k a}(g_n-\vep_zf_n)h'_n
\nonumber \\
&&+{2n^2\over (\k a)^4}\Big[{1+\vep_t\over (\k_0a)^2}-{2\vep_t\over (\k a)^2}\Big]\Bigg\},
\nonumber \\
P^{\rm out}_z&=& {\b\over 4k_0}\Bigg\{-{1\over \k_0a}(2g_n+h_n-\vep_zf_n)g'_n
\nonumber \\
&&-{2n^2\over (\k_0a)^4}\Big[{2\over (\k_0a)^2}-{1+\vep_t\over (\k a)^2}\Big]\Bigg\}. 
\label{Pz-hybrid}
\eea
Here $f'_n$, $g'_n$, and $h'_n$ are the derivatives of $f_n$, $g_n$, and $h_n$, respectively.
In this derivation, we assume $\vep_t$ and $\vep_z$ are real, thus the arguments of the Bessel
functions are all real. We set $A=\sqrt{g_n+h_n}/(k_0aJ_n)$.
Use has also been made of the following integrals which can not be found 
in any mathematics manual
\bea
\int_0^x\big(J'^2_{n}(x)+{n^2\over x^2}J_n^2(x)\big)xdx&=&-{x^3\over 2}J^2_n(x)f'_n(x),
\nonumber \\
\int_0^x\big(I'^2_{n}(x)+{n^2\over x^2}I^2_n(x)\big)xdx&=&-{x^3\over 2}I^2_n(x)h'_n(x),
\nonumber \\
\int_x^\infty\big(K'^2_{n}(x)+{n^2\over x^2}K^2_n(x)\big)xdx&=&-{x^3\over 2}K^2_n(x)g'_n(x).
\eea
Explicitly, one has
\bea
f'_n(x)&=&-xf^2_n(x)-{2\over x}f_n(x)+{n^2\over x^3}-{1\over x},\nonumber \\
h'_n(x)&=&-xh^2_n(x)-{2\over x}h_n(x)+{n^2\over x^3}+{1\over x},\nonumber \\
g'_n(x)&=&xg^2_n(x)-{2\over x}g_n(x)-{n^2\over x^3}-{1\over x}.
\eea
We point out that the above expressions for $P_z$ can be readily modified for dielectric or
metallic cylindrical waveguide with the exchange of 
$f_n(ix)=-h_n(x)$ and $ixf'_n(ix)=-xh'_n(x)$.

\begin{figure}[htbp]
\center{
\includegraphics [angle=0, width=8.2cm]{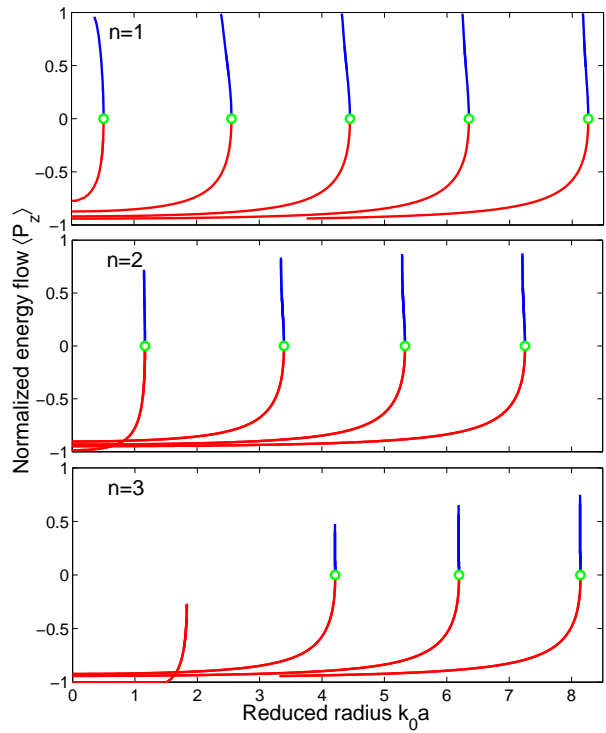}}
\caption{(Color online) Normalized energy flow $\langle P_z\rangle$ 
for the first few bands of the hybrid modes with $n=1,2,3$ on
an anisotropic waveguide with $\vep_t=-3$ and $\vep_z=2$. Open circles denote the degeneracy points where
$\langle P_z\rangle=0$.}
\label{fig-flow-hybrid}
\end{figure}

The normalized energy flow on a waveguide with $\vep_t=-3$ and $\vep_z=2$
for the hybrid modes with $n=1,2,3$ is plotted in Fig. \ref{fig-flow-hybrid}.

\subsubsection{Backward-wave and forward-wave hybrid modes}
The hybrid modes have similar features as the TM modes that both
forward waves and backward waves can co-exist within the same band.
For $k_0a\to 0$, $\k\simeq \k_0$, the hybrid modes have $d(k_0a)/d(\k_0a)<0$
and are all backward waves.
Only the modes near
the light line can be forward waves.
Making use of the expressions for $P_z$ in Eq. (\ref{Pz-hybrid}), 
one can prove that 
$d(k_0a)/d(\k_0a)\geq 0$ leads $P_z\geq0$ and vice versa.
The degeneracy of forward- and backward-wave modes is located at
$d(k_0a)/d(\k_0a)=0$ or $\langle P_z\rangle=0$.

For the hybrid modes with $n=1,2$, the modes near the light line are forward
waves since $d(k_0a)/d(\k_0a)>0$.
However as can be seen in Fig. \ref{fig-flow-hybrid},
the whole first band of the hybrid modes with $n=3$ are backward waves. 
Hybrid modes with higher $n$ will have more bands to be all backward waves.
If one further increases the angular index $n$ of the hybrid modes, 
more hybrid mode bands will be all backward waves.

\section{Slow and trapped light}

Recently Tsakmakidis {\it et al} \cite{Tsakmakidis} proposed to trap light in 
a tapered waveguide with double negative index.
The indefinite index waveguide we have studied so far in this paper 
can also be used to slow down and trap light. 
These waveguides can thus be used as optical buffers \cite{XiaF}.
The reason is that unlike the ordinary optical fiber, these waveguides support
both forward and backward waves.

For the anisotropic waveguide we have considered, $\vep_t<0$, 
one has $P^{\rm in}_z<0$ and $P^{\rm out}_z>0$ if one sets $\b>0$.
If $P_z=P^{\rm in}_z+P^{\rm out}_z<0$, the mode is a backward mode 
since the total energy flow is 
opposite to the phase velocity. Otherwise, the mode is a forward mode.
At the critical radius $a_c$, the backward and forward modes become degenerate,
the energy flow inside the waveguide cancels 
out that in the air. 
One can prove that at the critical radius $a_c$ where $P_z=0$,
the group velocity is indeed zero. One does not need to
know the material dispersion to locate the zero group velocity point.
This is due to the fact that for these waveguides, 
the dispersion due to geometric confinement dominates the material dispersion at 
and around the critical radius.

\begin{figure}[htbp]
\center{
\includegraphics [angle=0, width=7cm]{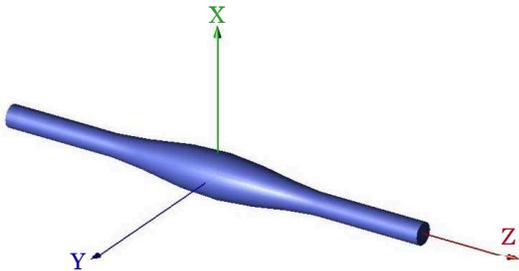}}
\caption{(Color online) A sketch of the slow light waveguide.}
\label{fig-sketch-slowlight}
\end{figure}

The unique properties of the modes on anisotropic waveguide can be used to
slow down and even trap light.
Even though the waveguide supports infinite number of both TM and hybrid modes
at any fixed radius and frequency, with appropriate laser coupling,
the excitation of the hybrid modes in the waveguide can be suppressed or even eliminated. 
Among the TM modes, the first TM mode will be more favorably excited.
Furthermore, due to the material dissipation, 
the first TM mode will propagate the longest distance. 
The rest of the TM modes will all decay out at about half 
the decay length of the first TM mode.
It is the first TM band which can be used for slow light application.
Unlike the double negative waveguide \cite{Tsakmakidis}, 
the anisotropic waveguide will slow down and trap light if one increases the radius to the critical radius. 
A sketch of a slow light waveguide is shown in Fig. \ref{fig-sketch-slowlight}.

\section{Realization of extremely anisotropic nanowires}

\begin{figure}[htbp]
\center{
\includegraphics [angle=0, width=7cm]{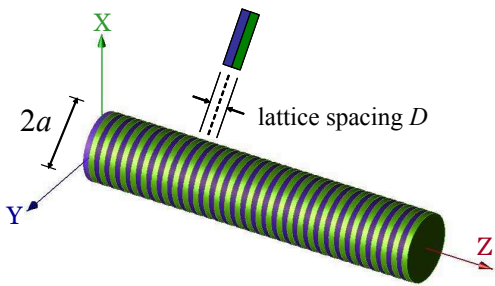}}
\caption{(Color online) A sketch of the realization of an anisotropic 
nanowire made of alternative disks of metal and dielectric.}
\label{fig-sketch-FDTD}
\end{figure}

These extremely anisotropic media can be realized in metamaterials.
According to the effective medium theory \cite{Sihvola,Lu07}, 
one has for multilayered structure of dielectric $\vep_a$ and metal $\vep_m$,
the effective permittivities are
\bea
\vep_t&=&f\vep_m+(1-f)\vep_a,
\nonumber \\
\vep_z&=&{\vep_a\vep_m\over f\vep_a+(1-f)\vep_m}.
\eea
Here $f$ is the filling ratio of the metal.
For $f>f_{\rm min}\equiv \vep_a/(\vep_a-\Re\vep_m)$, one has $\Re\vep_t<0$.
A realization of the anisotropic nanowire is shown in Fig. \ref{fig-sketch-FDTD}.

We first consider a metamaterial waveguide at a fixed wavelength.
For silver at $\l=488$ nm, one has $\vep_m=-9.121 + 0.304i$ \cite{Palik}.
A nanowire made of alernative disks of silver and glass ($\vep_a=2.25$) of equal thickness
will have $\vep_t=-3.436 + 0.152i$ and $\vep_z=5.971 + 0.065i$ by using
the effective medium theory.  Here the disk thickness is 10 nm for both materials.
For example if one sets $a=60$ nm, one has $k_0a=0.7725$. The first three TM modes will have 
$\b a=    2.2525 - 0.0747i, 4.9318 - 0.1419i, 7.4031 - 0.2088i$. 
Thus one has $\l_\b\sim 167$ nm and phase refractive index $n_{\rm eff}=2.92$ 
for the first TM mode. 
The decay length is 803, 423, and 287 nm, respectively.
After traveling about 420 nm along the nanowire, only the first one will survive.

\begin{figure}[htbp]
\center{
\includegraphics [angle=0, width=8.2cm]{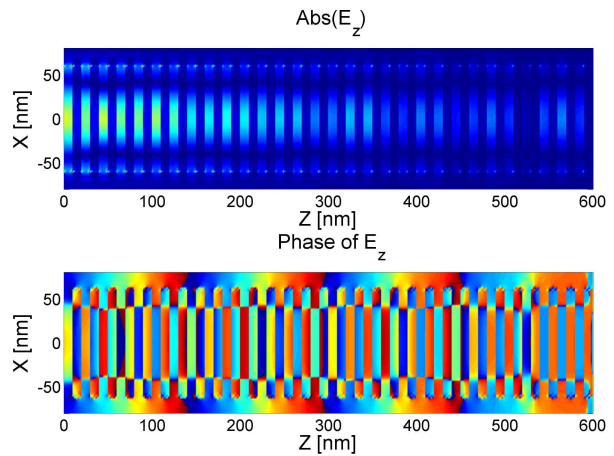}}
\caption{(Color online) FDTD simulation of the amplitude and phase propapation 
of the longitudinal electric field $E_z$ along the 
nanowire with radius $a=60$ nm at $\l=488$ nm. The metamaterial nanowire
consists of alternative disks of silver and glass disks of thickness 10 nm.}
\label{fig-FDTD-phase}
\end{figure}

Finite-difference time-domain (FDTD) simulations \cite{Taflove} were performed to
obtain the effective index $n_{\rm eff}$ of the metamaterial nanowire. 
The procedure is the following.
We illuminate the free-standing nanowire
of finite length with a Gaussian beam, then get $E_z$ after the termination of the simulation.
The length of the waveguide is set to be larger than the decay length of the first TM mode.
We get the phase from $E_z$, then determine $\b$. 
Though the waveguide supports infinite number of modes including TM and hybrid modes, 
our method is legitimate due to the following two reasons. 
First that the excitation of hybrid modes is small due to the 
profile of the incident Gaussion beam. So mainly the TM  modes are excited.
Second that due to the dissipation in the metal, after certain distance, only the first TM mode
will survive. Thus the phase propagation is mainly due to the first TM mode.
The amplitide and phase propagation of $E_z$ along the above metamaterial nanowire 
is shown in Fig. \ref{fig-FDTD-phase}. 

\begin{figure}[htbp]
\center{
\includegraphics [angle=0, width=7.5cm]{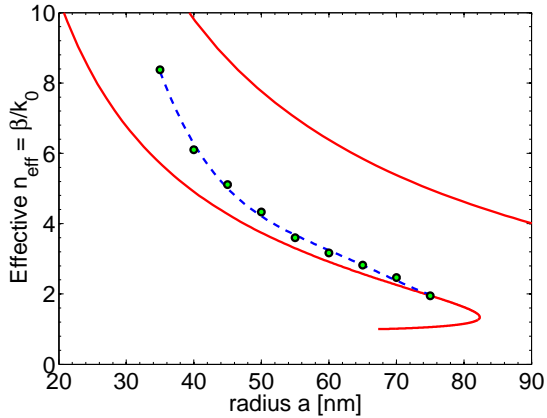}}
\caption{(Color online) The effective index $n_{\rm eff}$ of the first TM band on 
a nanowire with different radius at $\l=488$ nm. 
The nanowire is made of alternative disks of silver and (see Fig. \ref{fig-sketch-FDTD}). 
The disk thickness is 10 nm. 
Filled circle (green) is obtained from FDTD simulations. 
The dashed line (blue) is the fitting of simulation data.
The solid line (red) is calculated from band equation with 
effective index $\vep_t=-3.436 + 0.152i$ and $\vep_z=5.971 + 0.065i$.}
\label{fig-FDTD-488}
\end{figure}

The relation between the effective index $n_{\rm eff}$ and the nanowire radius $a$ 
is shown in Fig. \ref{fig-FDTD-488}. 
Very good agreement between FDTD simulations and analytical results has been obtained.
However for small radius, there is noticeable discrepancy.
This is expected since when the radius is comparable with the lattice spacing
of the multilayered metamaterial,
the effective medium theory will fail. 
We have also performed FDTD simulations for the nanowire with smaller lattice spacing.
Better agreement is indeed obtained.

\begin{figure}[htbp]
\center{
\includegraphics [angle=0, width=2.8in]{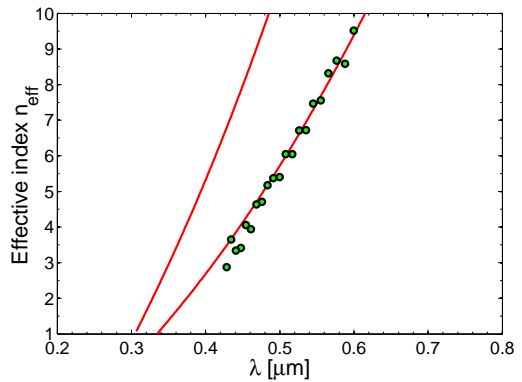}}
\caption{(Color online) The effective index $n_{\rm eff}$ for the TM modes on
a nanowire with radius $a=40$ nm. The nanowire is a stack of equally thick
alternative disks made of a Drude metal $\vep_m=1-k_p^2/k_0(k_0+i\Gamma)$ and glass.
Here $k_pa=1.64$ and $\Gamma a=0.0155$. The analytical curves (solid) are calculated 
by using real $\vep_m$.}
\label{fig-FDTD-fixed-a}
\end{figure}

We also consider the band structure for different frequencies.
The permittivity given by Eq. (\ref{ep-metal-dielec}) can be realized 
through the multilayered heterostructure with Drude metal with $\vep_m=1-k_p^2/k_0$ 
and dielectric $\vep_a$.
The nanowire is made of alternative disks of a Drude metal and a dielectric. 
The band structure and the effective index $n_{\rm eff}$ of the TM modes
are shown in Fig. \ref{fig-drude-band} and Fig. \ref{fig-drude-neff}, respectively.
One noticeable feature of these bands is the flatness of each band,
which indicates small group velocity.
We have erformed the FDTD simulation for different frequencies 
for nanowire with a fixed radius. 
The results are shown in Fig. \ref{fig-FDTD-fixed-a}.
Again good agreement between FDTD simulations and analytical results is achieved.

\section{conclusions}

Indefinite index materials can be used to achieve negative refraction \cite{Hoffman} 
and hyperlensing \cite{LiuZ,Smolyaninov}. 
They can also be used as superlens \cite{Lu07}. 
In this paper, we consider the wave propagation along a cylindrical waveguide with
anisotropic optical constant.
We have derived the eigenmodes equation and obtained the solutions for all the propagation modes.
The field profiles and the energy flow on the waveguide are also analyzed.
Closed-form expressions for the energy flow
for all the modes are derived.
For extremely anisotropic cylinder where the transverse component of the permittivity 
is negative and the longitudinal is positive ($\vep_t<0$, $\vep_z>0$), the waveguide supports TM 
and hybrid modes but not the TE modes. 
Among the supported TM modes, at most only one mode can be forward wave.
The rest of them are backward waves.

The case that $\vep_t>0$ and $\vep_z<0$ 
can be discussed similarly.
Anisotropic waveguides of cross section other than circle were also considered. 
The results will be published elsewhere.

Possible realization of these extremely anisotropic nanowires are proposed. 
Extensive FDTD simulations have been performed and confirmed our analytical results.

Two unique properties have been revealed for the modes on nanowire waveguides
made of indefinite index metamaterials.
The first is that the backward-wave modes can have very large effective
index. These nanowires can be used as phase shifters and filters in optics 
and telecommunication.
The second is that the waveguide supports modes of zero group velocity.
This is due to the fact that the waveguide can support both forward and backward waves at
a fixed radius. If the waveguide is tapered, at certain critical radius,
the two modes will be degenerate and carry zero net energy flow.
At other radii, these waveguides support modes with small group velocity.
These waveguides can also be used as ultra-compact optical buffer \cite{XiaF} 
in integrated optical circuits.

\begin{acknowledgments}
This work was supported by the Air Force Research Laboratories, Hanscom
through FA8718-06-C-0045 and the National Science Foundation
through PHY-0457002.
\end{acknowledgments}

\appendix
\section{Pad\'e approximant of $f_n(x)$}
Consider the function
$f_n(x)=J'_n(x)/[x J_n(x)]$.
Let $x_{n,m}$ be the $m$-th zero of the Bessel function $J_n(x)$.
We also denote $x'_{n,m}$ as the zeros of $J'_n(x)$.
In order to get the eigen modes on the anisotropic waveguide easily, we may need 
 the inverse function $f^{-1}_n(x)$ in the interval $[x_{n,m},x'_{n,m}]$.
We consider the Pad\'e approximant to the function $f_n(x)$,
\be
f_n(x)\simeq {(x-x'_{n,m})(x-x_{n,m}-b)\over c(x-x_{n,m})(x-x_{n,m}-a)}.\label{pade-fn}
\ee
Instead of fixing the three unknowns through the coefficients of the 
Taylor expansion of $f_n(x)$,
here we determine them by the exact values of $f_n(x)$ at some points.
Since there are three unknowns, we only need the value of $f_n(x)$ at three 
points. For simplicity, we evaluate $f_n(x)$ at three evenly spaced points
\be
f_{n,j}\equiv f_n(x_{n,m}+j\Delta/4)
\ee
for $j=1,2,3$. Here $\Delta=x'_{n,m}-x_{n,m}$. One thus has
$f_{n,j}=(4-j)(j\D-4b)/[jc(j\D-4a)$.
We further define
\be
\s_1=3f_{n,2}/f_{n,1},\quad \s_2=3f_{n,3}/f_{n,2}.
\ee
After some manipulations of the algebra, one obtains
\bea
a&=&{\D\over 4}\Big(2+{\s_2-\s_1^{-1}\over \s_2+\s_1^{-1}-2}\Big),\nonumber \\
b&=&{\D\over 4}\Big(2-{\s_1-\s_2^{-1}\over \s_1+\s_2^{-1}-2}\Big),\nonumber \\
c&=&-{1\over f_2}{\D-2b\over \D-2a}.
\eea
The inverse of the function $f_n(x)$ can be obtained as one of the roots
of a quadratic equation. As it turns out,
the expression obtained in this way gives very good approximation to $f^{-1}_n(x)$.

Once the inverse function is obtained, one has $Ka=f^{-1}_0(y)$ with 
$y=g_0(\k_0a)/\vep_z$ for the TM modes. 
Together with the asymptotic expressions of $g_n(x)$ and $h_n(x)$, Eq. (\ref{pade-fn})
can also be used to obtain approximate solutions of the hybrid modes.

\end{document}